\documentclass[pra,showpacs,12pt,tightenlines]{revtex4b5}

\newcommand{\ket}[1]{|#1\rangle}
\newcommand{\bra}[1]{\langle #1 |}
\newcommand{\mymid}{\,:\,}
\newcommand{\qed}{\hspace*{\fill}\rule{1.3ex}{1.3ex}}

\begin{document}
\preprint{quant-ph/0011047}
\title{Fidelity of a $t$-error correcting\\
quantum code with more than $t$ errors}
\author{Ryutaroh Matsumoto}
\email{ryutaroh@rmatsumoto.org}
\homepage{http://www.rmatsumoto.org/}
\thanks{%
To be published in Phys.\ Rev.\ A.}
\affiliation{Department of Communications and Integrated Systems,
Tokyo Institute of Technology, 152-8552 Japan}
\date{April 25, 2001}
\begin{abstract}
It is important to study the behavior of a $t$-error correcting
quantum code when the number of errors is greater than $t$,
because it is likely that there are also small errors
besides $t$ large correctable errors.
We estimate the fidelity of
a $t$-error correcting stabilizer code over
a general memoryless channel, allowing more than
$t$ errors.
We also show that the fidelity can be made arbitrary close
to $1$ by increasing the code length.
\end{abstract}
\pacs{03.67.Hk, 89.70.+c}
\maketitle
\section{Introduction}
In the study of the quantum error-correcting codes,
it is usually assumed that only a small number of qubits are affected
and the rest of qubits are left unchanged.
However, it is important to study the behavior of a $t$-error correcting
quantum code when the number of errors is greater than $t$,
because it is likely that there are also \emph{small} errors
besides $t$ large correctable errors.
The goal of this paper is to provide a lower bound for the fidelity
of the quantum error correction under the general noise
model without any approximation.
The fact that quantum error-correcting codes work under the general
noise model seems a folklore result,
and the original contributions of this paper are
a rigorous 
proof and a quantitative relation between the fidelity and the noisiness 
of the channel.

The following researches have been done prior to this paper.
It has been informally argued in Ref.\ \cite[Sec.\  VI]{calderbank96}
that those small errors
do not result in a large error in the recovered quantum state.
The first rigorous analysis was done in
Ref.\ \cite[Sec.\  5.4]{knill97}, in which
the authors assumed that the channel was memoryless,
that is, each qubit interacts with different environment,
and there was a
scalar multiple of the identity operator in an operator sum
representation of the channel superoperator.
In Ref.\ \cite[Sec.\  7.4.2]{preskill99},
quantitative relations between the fidelity and the noisiness
were given for two specific classes of memoryless channels.
Aharonov and Ben-Or \cite[Sec.\  8]{aharonov99}
analyzed the fault-tolerant quantum computation
under the general noise model that is equivalent to a memoryless channel,
and showed that if the channel is not too noisy then the error-free
computation is possible.
However, they did not provide a quantitative condition of general channels
allowing the error-free computation. (They provided that of restricted
channels.)

In this paper we assume that a unitary representation of the
channel superoperator has large identity component (Assumption
3), and
we give a lower bound (Eq.\ (\ref{lastevaluation})) for the average of the fidelity between the original state and
the recovered state without using any approximation,
where the average is taken over the measurement outcome
in the error correction process.
As a consequence
we show that the average of the fidelity can be made arbitrary close
to $1$ over a general memoryless channel by increasing the code length.
This fact has been proved only over specific classes of quantum channels
\cite{knill97,preskill99}.
Our estimation is restricted to the stabilizer quantum codes
introduced in Refs.\ \cite{calderbank97,calderbank98,gottesman96},
which include almost all good quantum codes discovered so far.
It should be noted that the essential idea in our analysis already
appeared in Ref.\ \cite[Sec.\  7.4]{preskill99}.

This paper is organized as follows:
In Sec.\  \ref{sec:b} we introduce notations used in this paper,
and review the stabilizer quantum codes and their
error correction process.
In Sec.\  \ref{sec:c} we give a lower bound for the fidelity.
In Sec.\  \ref{sec:d} several consequences and generalizations are
discussed.

\section{Preliminaries}\label{sec:b}
\subsection{Notations}\label{sec:notation}
Let $\mathcal{H}$ be a Hilbert space.
We denote by $\mathcal{S}(\mathcal{H})$
the set of density operators on $\mathcal{H}$.
For a density operator $\rho$ on $\mathcal{H}$ and
a state vector $\ket{\psi}\in \mathcal{H}$,
the fidelity \cite{jozsa94,uhlmann76} between them is defined by
\[
F(\ket{\psi}, \rho) = \bra{\psi}\rho\ket{\psi}.
\]
It measures how close $\ket{\psi}$ and $\rho$ are.

In this paper we consider $t$-error correcting
$[[n,k]]$ binary quantum codes unless otherwise stated.
Let $H_2$ be the Hilbert space of dimension $2$.
Let $\Gamma$ be a superoperator on $H_2$,
that is,
a trace-preserving completely positive linear map from
$\mathcal{S}(H_2)$ to $\mathcal{S}(H_2)$.
We assume that the channel is represented by $\Gamma$,
which means that
when we send a density operator $\rho \in \mathcal{S}(H_2)$ through the channel
we get $\Gamma(\rho) \in \mathcal{S}(H_2)$ at the receiving end.
The channel considered in this paper is assumed to be memoryless.
So when we send a state $\rho \in \mathcal{S}(H_2^{\otimes n})$
we get $\Gamma^{\otimes n}(\rho)$.

We shall review the unitary representation of a superoperator
\cite{book:kraus}.
A simplified proof can be found in Ref.\ \cite[Appendix]{schumacher96}.
Let $\Gamma$ be a superoperator on a Hilbert space $\mathcal{H}$.
Then there exist a Hilbert space $\mathcal{H}_E$,
a state vector $\ket{0_E}\in \mathcal{H}_E$,
and a unitary operator $U$ on $\mathcal{H}\otimes \mathcal{H}_E$
such that
\begin{equation}
\Gamma(\rho) = \mathrm{Tr}_E(U(\rho\otimes |0_E\rangle \langle0_E|)U^*),
\label{eq:unitary}
\end{equation}
for all $\rho\in\mathcal{S}(\mathcal{H})$,
where $\mathrm{Tr}_E$ is the partial trace over $\mathcal{H}_E$.
That is called a unitary representation of $\Gamma$.

\subsection{Stabilizer quantum codes}\label{sec:stabilizer}
In this subsection we review the method of quantum error correction
proposed in Refs.\ \cite{calderbank97,calderbank98,gottesman96}.
Let
\[
\sigma_x = \left(\begin{array}{cc}0&1\\ 1&0\end{array}\right),\;
\sigma_z = \left(\begin{array}{cc}1&0\\ 0&-1\end{array}\right),
\]
and $E = \{ \pm w_1 \otimes $ $\cdots$ $\otimes w_n\}$,
where $w_i$ is either $I$, $\sigma_x$, $\sigma_z$ or $\sigma_x\sigma_z$.
The set $E$ is a noncommutative group with matrix multiplication as
its group operation.
Let $S$ be a commutative subgroup of $E$.
A quantum error-correcting code $Q \subset H_2^{\otimes n}$ is defined as
an eigenspace of $S$.

For $M\in E$ we define $MQ = \{M\ket{\varphi} \mymid
\ket{\varphi} \in Q\}$. The set $MQ$ is also an eigenspace of $S$
for any $M\in E$.
Moreover $\{ MQ \mymid M \in E \}$ is equal to the set of
eigenspaces of $S$.
It follows that every eigenspace of $S$ has the same dimension.
Let $\dim Q = 2^k$. Then there are $2^{n-k}$ eigenspaces of $S$.
Let $S' = \{ N \in E \mymid MN = NM$ for all $M\in S\}$.
It is known that
\begin{equation}
S' = \{ M\in E\mymid MQ = Q \}. \label{eq:sequal}
\end{equation}

We shall describe the error correction procedure.
Let $H_\mathrm{env}$ be the Hilbert space representing the
environment around the channel.
Suppose that we send a pure state $\ket{\varphi}\in Q$,
and the environment is initially in a pure state $\ket{0_\mathrm{env}}
\in H_\mathrm{env}$.
Suppose also that we receive an entangled state $\ket{\psi} \in
H_2^{\otimes n}\otimes H_\mathrm{env}$.
We measure an observable of $H_2^{\otimes n}$ whose eigenspaces
are the same as those of $S$.
Then the state $\ket{\psi}$ is projected to
$\ket{\psi'} \in Q' \otimes H_\mathrm{env}$, where
$Q'$ is some eigenspace of $S$.

We will define the weight of an operator $M \in E$ for error
correction. Let $M = \pm w_1 \otimes \cdots \otimes w_n$,
where $w_i$ is either $I$, $\sigma_x$, $\sigma_z$ or $\sigma_x \sigma_z$.
The weight of $M$ is defined to be
$\sharp \{ i \mymid w_i \neq I\}$, and
denoted by $w(M)$.
We define the numbers $d$ and $d'$ by
\begin{eqnarray*}
d &=& \min\{w(M) \mymid M\in S' \mbox{ and } \pm M \notin S \},\\
d' &=& \min\{ w(M) \mymid M \in S' \mbox{ and } \pm M \neq I\}.
\end{eqnarray*}
The number $d$ is called the minimum distance of $Q$.
The code $Q$ is said to be pure if $d = d'$ and
impure if $d > d'$.
We define $t = \lfloor (d-1)/2 \rfloor$.

There are many operators $M \in E$ such that $MQ = Q'$.
Let $M$ be an operator whose weight is minimum among them.
Note that if the weight of $M$ is greater than $\lfloor
(d'-1)/2\rfloor$ then
there may be another operator $M'$ such that $w(M') = w(M)$,
$M'Q = Q'$ and $M \neq \pm M'$.
We guess that the original pure state is
$(M^{-1}\otimes I_{\mathrm{env}}) \ket{\psi'}$,
where $I_\mathrm{env}$ is the identity operator on $H_\mathrm{env}$.

If the number of errors $\leq t$,
then $\ket{\psi'}$ is the tensor product of
$\ket{\varphi'} \in Q'$ and some pure state in $H_\mathrm{env}$,
and $M^{-1} \ket{\varphi'} = \ket{\varphi}$.
However, we do not make such assumption,
and we shall analyze the closeness (fidelity)
between $\ket{\varphi}$ and $M^{-1} \mathrm{Tr}_{H_\mathrm{env}}
(\ket{\psi'}\bra{\psi'}) (M^{-1})^*$.

We shall use the following fact later.

\noindent\textbf{Proposition 1}\ \itshape
Let $M' \in E$ be an operator such that
$M'Q = MQ$.
If $\pm M' \notin MS$ then $w(M') > t$,
where $MS = \{ MN \mymid N\in S\}$.
\upshape

\noindent\emph{Proof.}
If $w(M) > t$ then $w(M') > t$ by the definition of $M$.
Suppose that $w(M) \leq t$ and $w(M') \leq t$.
Then $w(M^{-1} M') \leq 2t < d$, $M^{-1} M' \in S'$
by Eq.\ (\ref{eq:sequal}),
and $M^{-1}M' \notin S$.
This contradicts to the definition of $d$.
\qed

\section{Lower bound for the fidelity}\label{sec:c}
In this section we consider
the fidelity between the original state and recovered state.
Let $\Gamma$ be the channel superoperator of $H_2$ as in
Sec.\  \ref{sec:notation}.
Since $I$, $\sigma_x$, $\sigma_z$ and $\sigma_x \sigma_z$
form a basis of linear operators on $H_2$,
in a unitary representation of $\Gamma$
we can write $U$ in Eq.\ (\ref{eq:unitary}) as
\[
I \otimes L_{0,0} + \sigma_x \otimes L_{1,0} +
\sigma_z \otimes L_{0,1} + \sigma_x \sigma_z L_{1,1},
\]
where
$L_{i,j}$ is a linear operator on
a Hilbert space $H_E$.
Let $\ket{0_E}$ be the initial state of $H_E$.

\noindent\textbf{Lemma 2}\ \itshape
We retain notations as above.
$\| L_{0,0} 0_E\| \leq 1$,
where $\| \cdot\|$ denotes the norm of a vector $\cdot$.
\upshape

\noindent\emph{Proof.}
Let $\{\ket{0}, \ket{1}\}$ be the orthonormal basis
such that $\sigma_x \ket{0} = \ket{1}$,
$\sigma_x \ket{1} = \ket{0}$,
$\sigma_z \ket{0} = \ket{0}$,
and $\sigma_z \ket{1} = -\ket{1}$.
Then we have
\begin{eqnarray*}
U(\ket{0}\otimes\ket{0_E}) &=&
\ket{0} \otimes (L_{0,0}\ket{0_E} + L_{0,1}\ket{0_E}) +
\ket{1} \otimes (L_{1,0}\ket{0_E} + L_{1,1}\ket{0_E},\\
U(\ket{1}\otimes\ket{0_E}) &=&
\ket{1} \otimes (L_{0,0}\ket{0_E} - L_{0,1}\ket{0_E}) +
\ket{0} \otimes (L_{1,0}\ket{0_E} - L_{1,1}\ket{0_E}.
\end{eqnarray*}
Since both vectors are of unit norm,
it follows that
\begin{eqnarray*}
\| L_{0,0}\ket{0_E} + L_{0,1}\ket{0_E} \| & \leq & 1,\\
\| L_{0,0}\ket{0_E} - L_{0,1}\ket{0_E} \| & \leq & 1.
\end{eqnarray*}
We conclude $\| L_{0,0}0_E\| \leq 1$.
\qed

\noindent\textbf{Assumption 3}\ \itshape
Assume that
\[
\| L_{0,1} 0_E \|^2 + \| L_{1,0} 0_E \|^2  + \| L_{1,1} 0_E \|^2 =  p.
\]
\upshape

Hereafter we denote $H_E^{\otimes n}$ by $H_\mathrm{env}$ and
$\ket{0_E}\otimes \cdots \otimes \ket{0_E}$ by
$\ket{0_\mathrm{env}}$.
Suppose that we send $\ket{\varphi} \in Q$ and
the recovered state is $M^{-1} \mathrm{Tr}_{H_\mathrm{env}}
(\ket{\psi'}\bra{\psi'})(M^{-1})^*$
as in Sec.\  \ref{sec:stabilizer}.

We shall consider the average of $F(\ket{\varphi},
M^{-1} \mathrm{Tr}_{H_\mathrm{env}}
(\ket{\psi'}\bra{\psi'})(M^{-1})^*)$ for an arbitrary fixed state
$\ket{\varphi}\in Q$ under the assumption that
the channel is memoryless. The superoperator of the channel is
$\Gamma^{\otimes n}$.
Let $\mathbf{Z}_2 = \{0, 1\}$ with the addition and the multiplication
taken modulo $2$. For $\vec{a} = (a_1$, \ldots, $a_n) \in \mathbf{Z}_2^n$,
we define
\begin{eqnarray*}
X(\vec{a}) &=& \sigma_x^{a_1} \otimes \cdots \otimes \sigma_x^{a_n},\\
Z(\vec{a}) &=& \sigma_z^{a_1} \otimes \cdots \otimes \sigma_z^{a_n}.
\end{eqnarray*}
Then a unitary representation of $\Gamma^{\otimes n}$ can be written as
\[
\sum_{\vec{a}, \vec{b} \in \mathbf{Z}_2^n}
 X(\vec{a})Z(\vec{b}) \otimes L_{\vec{a}\vec{b}},
\]
where
\[
L_{\vec{a}\vec{b}} = L_{a_1,b_1} \otimes \cdots \otimes L_{a_n,b_n}.
\]
Let $\ket{\psi}$ be as in Sec.\  \ref{sec:stabilizer}.
By notations defined so far, $\ket{\psi}$ can be written as
\[
\ket{\psi}
=
\sum_{\vec{a},\vec{b} \in \mathbf{Z}_2^n}
X(\vec{a})Z(\vec{b}) \ket{\varphi}
\otimes L_{\vec{a}\vec{b}} \ket{0_\mathrm{env}}.
\]

Let $Q'$ be an eigenspace of $S$.
We shall consider the probability $P_{Q'}$ of $\ket{\psi}$ being projected
to $Q' \otimes H_\mathrm{env}$
after the measurement.
Let $(\vec{a}_{Q'}$, $\vec{b}_{Q'})$ be a pair of vectors
such that $X(\vec{a}_{Q'})Z(\vec{b}_{Q'})Q = Q'$ and that
if $M'Q = Q'$ then $w(M') \geq w(X(\vec{a}_{Q'})Z(\vec{b}_{Q'}))$.
Observe that if $w(X(\vec{a}_{Q'})Z(\vec{b}_{Q'})) > \lfloor (d'-1)/2
\rfloor$
then there may be another operator
$M'$ such that $M'Q = Q'$, $w(M')
= w(X(\vec{a}_{Q'})Z(\vec{b}_{Q'}))$
and $M' \neq \pm X(\vec{a}_{Q'})Z(\vec{b}_{Q'})$.
This implies that $(\vec{a}_{Q'}$, $\vec{b}_{Q'})$ is not uniquely
determined by $Q'$.
One may choose whichever $(\vec{a}_{Q'}$, $\vec{b}_{Q'})$
provided that $X(\vec{a}_{Q'})Z(\vec{b}_{Q'})$ has the minimum weight
(see also Sec.\  \ref{sec:bounded}).
Let $T_{Q'} = \{ (\vec{c}, \vec{d}) \in \mathbf{Z}_2^n \times
\mathbf{Z}_2^n \mymid X(\vec{c})Z(\vec{d}) Q = Q'$
and $\pm X(\vec{c})Z(\vec{d}) \notin X(\vec{a}_{Q'})Z(\vec{b}_{Q'})S \}$.
We define
\begin{eqnarray*}
\ket{\sigma_{Q'}} &=& \sum_{\renewcommand{\arraystretch}{0}\begin{array}{c}\scriptstyle \vec{c},\vec{d}\in\mathbf{Z}_2^n\\
\scriptstyle \pm X(\vec{c}) Z(\vec{d}) \in X(\vec{a}_{Q'})Z(\vec{b}_{Q'})
S \end{array}}
X(\vec{c})Z(\vec{d}) \ket{\varphi}
\otimes L_{\vec{c}\vec{d}} \ket{0_\mathrm{env}},\\
\ket{\sigma'_{Q'}} &=&
\sum_{(\vec{c}_{Q'},\vec{d}_{Q'}) \in T_{Q'}}
X(\vec{c}_{Q'})Z(\vec{d}_{Q'}) \ket{\varphi}
\otimes L_{\vec{c}_{Q'}\vec{d}_{Q'}} \ket{0_\mathrm{env}}.
\end{eqnarray*}
Observe that
\begin{equation}
\ket{\psi} = \sum_{Q'\textnormal{\scriptsize\ is an eigenspace of }S}
\ket{\sigma_{Q'}+\sigma'_{Q'}}, \label{sumeigenvector}
\end{equation}
and the projection of $\ket{\psi}$ to $Q' \otimes H_\mathrm{env}$ is
$\ket{\sigma_{Q'}+\sigma'_{Q'}}$.
Thus $P_{Q'}$ is given by $\| \sigma_{Q'}+\sigma'_{Q'} \|^2$.

Let $\ket{\psi'} = \ket{\sigma_{Q'}+\sigma'_{Q'}} / \| \sigma_{Q'}+\sigma'_{Q'} \|$, and $M = X(\vec{a}_{Q'})Z(\vec{b}_{Q'})$.
Next we shall calculate a lower bound for the fidelity between
$\ket{\varphi}$ and the recovered state
$M^{-1} \mathrm{Tr}_{H_\mathrm{env}} (\ket{\psi'}\bra{\psi'})
(M^{-1})^*$
when $\ket{\psi}$ is projected to $\ket{\sigma_{Q'}+\sigma'_{Q'}} \in Q' \otimes H_\mathrm{env}$
after the measurement.
Observe that taking partial trace over $H_\mathrm{env}$
and applying $M^{-1}$ to $\ket{\sigma_{Q'}}$ and $\ket{\sigma_{Q'}+
\sigma'_{Q'}}$ yields the original state $\ket{\varphi}$ and the recovered state
$M^{-1} \mathrm{Tr}_{H_\mathrm{env}} (\ket{\psi'}\bra{\psi'})
(M^{-1})^*$, respectively.
The fidelity between
$\ket{\varphi}$ and
$M^{-1} \mathrm{Tr}_{H_\mathrm{env}} (\ket{\psi'}\bra{\psi'})
(M^{-1})^*$ is not less than
that between $\ket{\sigma_{Q'}}$
and 
$\ket{\sigma_{Q'}+\sigma'_{Q'}}$,
because the fidelity does not decrease by unitary operations and
taking partial trace
\cite{jozsa94b}.
We shall calculate a lower bound for the fidelity $F_{Q'}$ between
$\ket{\sigma_{Q'}}$
and 
$\ket{\sigma_{Q'}+\sigma'_{Q'}}$.
\begin{eqnarray*}
1-F_{Q'} &=& 1 - \frac{\langle\sigma_{Q'}|\sigma_{Q'}+\sigma'_{Q'}\rangle\langle
\sigma_{Q'}+\sigma'_{Q'}|\sigma_{Q'}\rangle}%
{\langle\sigma_{Q'}|\sigma_{Q'}\rangle\langle\sigma_{Q'}+\sigma'_{Q'}|
\sigma_{Q'}+\sigma'_{Q'}\rangle} \\
&=& \frac{\langle\sigma_{Q'}|\sigma_{Q'}\rangle\langle\sigma'_{Q'}|\sigma'_{Q'}\rangle
-\langle\sigma'_{Q'}|\sigma_{Q'}\rangle\langle\sigma_{Q'}|\sigma'_{Q'}\rangle}%
{\langle\sigma_{Q'}|\sigma_{Q'}\rangle\langle\sigma_{Q'}+\sigma'_{Q'}|
\sigma_{Q'}+\sigma'_{Q'}\rangle} \\
&\leq &\frac{\langle\sigma_{Q'}|\sigma_{Q'}\rangle\langle\sigma'_{Q'}|\sigma'_{Q'}\rangle}%
{\langle\sigma_{Q'}|\sigma_{Q'}\rangle\langle\sigma_{Q'}+\sigma'_{Q'}|
\sigma_{Q'}+\sigma'_{Q'}\rangle} \\
&=& \frac{\langle\sigma'_{Q'}|\sigma'_{Q'}\rangle}%
{\langle\sigma_{Q'}+\sigma'_{Q'}|
\sigma_{Q'}+\sigma'_{Q'}\rangle}.
\end{eqnarray*}

We shall calculate a lower bound for the average of $1-F_{Q'}$,
where the average is taken over the measurement outcome.
The following fact will be used.
For a pair of vectors $(\vec{a}, \vec{b})$,
$w(\vec{a}, \vec{b})$ denotes $w(X(\vec{a})Z(\vec{b}))$.

\noindent\textbf{Proposition 4}
\[
\bigcup_{Q'\textnormal{\scriptsize\ is an eigenspace of }S} T_{Q'}
\subset
\{ (\vec{a},\vec{b}) \in \mathbf{Z}_2^n \times \mathbf{Z}_2^n \mymid
w(\vec{a},\vec{b}) > t \}.
\]

\noindent\emph{Proof.}
The assertion follows from Proposition 1.
\qed

In the following calculation $Q'$ runs through the set of eigenspaces of $S$.
\begin{eqnarray}
\sum_{Q'} P_{Q'} (1-F_{Q'}) &\leq& \sum_{Q'} \langle\sigma'_{Q'}|\sigma'_{Q'}\rangle \nonumber\\
&=&  \sum_{Q'} \left\| \sum_{(\vec{c}_{Q'},\vec{d}_{Q'}) \in T_{Q'}}
X(\vec{c}_{Q'})Z(\vec{d}_{Q'}) \ket{\varphi}
\otimes L_{\vec{c}_{Q'}\vec{d}_{Q'}} \ket{0_\mathrm{env}}\right\|^2\nonumber\\
&\leq&  \sum_{Q'} \sum_{(\vec{c}_{Q'},\vec{d}_{Q'}) \in T_{Q'}}
\|X(\vec{c}_{Q'})Z(\vec{d}_{Q'}) \ket{\varphi}
\otimes L_{\vec{c}_{Q'}\vec{d}_{Q'}} \ket{0_\mathrm{env}}\|^2\nonumber\\
&\leq&
\sum_{\renewcommand{\arraystretch}{0}\begin{array}{c}\scriptstyle \vec{c},\vec{d}\in\mathbf{Z}_2^n\\
\scriptstyle w(\vec{c},\vec{d})>t\end{array}}
\|X(\vec{c})Z(\vec{d}) \ket{\varphi}
\otimes L_{\vec{c}\vec{d}} \ket{0_\mathrm{env}}\|^2 \mbox{ (by Proposition 4)} \nonumber\\
&=&  \sum_{\renewcommand{\arraystretch}{0}\begin{array}{c}\scriptstyle \vec{c},\vec{d}\in\mathbf{Z}_2^n\\
\scriptstyle w(\vec{c},\vec{d})>t\end{array}} \| L_{\vec{c}\vec{d}} 0_\mathrm{env}\|^2. \label{eq:cd}
\end{eqnarray}
For a vector $\vec{a} = (a_1$, \ldots, $a_n) \in \mathbf{Z}_2^n$,
let
\begin{eqnarray*}
\ell(0) &=& \| L_{0,0} 0_E \|^2,\\
\ell(1) &=& \| L_{0,1} 0_E \|^2 + \| L_{1,0} 0_E \|^2 + \| L_{1,1} 0_E \|^2,\\
\Delta(\vec{a}) &=& \prod_{i=1}^n \ell(a_i),\\
h(\vec{a}) &=& \sharp \{ i \mymid a_i \neq 0\}.
\end{eqnarray*}
Observe that $\Delta(\vec{a}) \leq p^{h(\vec{a})}$ by
Assumption 3 and Lemma 2.
For vectors $\vec{a}$, $\vec{b} \in \mathbf{Z}_2^n$,
let $\mathrm{or}(\vec{a}$, $\vec{b})$ be the bitwise logical or
of them.
By these notations, for a vector $\vec{a} \in \mathbf{Z}_2^n$ we can see
\[
\sum_{\renewcommand{\arraystretch}{0}\begin{array}{c}\scriptstyle \vec{c},\vec{d}\in\mathbf{Z}_2^n\\
\scriptstyle \mathrm{or}(\vec{c},\vec{d})=\vec{a}\end{array}} \| L_{\vec{c}\vec{d}} 0_\mathrm{env}\|^2 = \Delta(\vec{a}) \leq p^{h(\vec{a})},
\]
and $w(\vec{c}$, $\vec{d}) = h(\mathrm{or}(\vec{c}$, $\vec{d}))$.
By these observations we can rewrite Eq.\ (\ref{eq:cd}) as
\begin{eqnarray}
\sum_{\renewcommand{\arraystretch}{0}\begin{array}{c}\scriptstyle \vec{c},\vec{d}\in\mathbf{Z}_2^n\\
\scriptstyle w(\vec{c},\vec{d})>t\end{array}} \| L_{\vec{c}\vec{d}} 0_\mathrm{env}\|^2 &=&
\sum_{\renewcommand{\arraystretch}{0}\begin{array}{c}\scriptstyle \vec{a}\in\mathbf{Z}_2^n\\
\scriptstyle h(\vec{a})>t\end{array}} \Delta(\vec{a})\label{beforelast}\\
&\leq&
\sum_{\renewcommand{\arraystretch}{0}\begin{array}{c}\scriptstyle \vec{a}\in\mathbf{Z}_2^n \\
\scriptstyle h(\vec{a})>t\end{array}} p^{h(\vec{a})} \label{usememoryless}\\
&=& \sum_{i=t+1}^n \left(\begin{array}{c}n\\ i\end{array}\right)
p^i. \nonumber
\end{eqnarray}

Thus
\begin{equation}
1- \sum_{i=t+1}^n \left(\begin{array}{c}n\\ i\end{array}\right)
p^i \label{lastevaluation}
\end{equation}
is a lower bound for the average of the fidelity between
the original state and the state recovered by a $t$-error
correcting quantum code of length $n$.

\noindent\textbf{Example 5}
By Ref.\ \cite[Table III]{calderbank98}
it is known that there exists a $[[25,5,7]]$ code.
We take it as an example.
Then we have $t=3$.
At $p=0.01$, the value of Eq.~(\ref{lastevaluation}) is $1-0.000132$,
and at $p=0.001$ the value of Eq.~(\ref{lastevaluation}) is $1-0.127\times 10^{-7}$.

\section{Consequences and generalizations}\label{sec:d}
\subsection{Error-free communication is asymptotically possible}\label{sec:asymptotic}
In classical information transmission
we can make the error probability arbitrary small by
increasing the code length.
The same result also holds in the quantum case.
Let $\alpha$ be a real number such that
$2p^\alpha < 1$.
Suppose that there exists a sequence of $t_i$-error correcting
quantum codes of length $n_i$ such that $t_i / n_i \rightarrow
\alpha$ and $n_i \rightarrow \infty$ as $i\rightarrow \infty$.
The existence of such a sequence is guaranteed by
the quantum Varshamov-Gilbert bound \cite[Theorem 2]{calderbank97}
in certain range of $\alpha$.

We shall consider the asymptotic behavior of Eq.\ (\ref{lastevaluation}):
\begin{eqnarray*}
\sum_{i=t+1}^n \left(\begin{array}{c}n\\ i\end{array}\right)
p^i &\leq& p^{t+1} \sum_{i=t+1}^n \left(\begin{array}{c}n\\ i\end{array}\right)\\
&\leq& p^{t+1} \sum_{i=1}^n \left(\begin{array}{c}n\\ i\end{array}\right)\\
&=& p^{t+1} 2^n.
\end{eqnarray*}
If $t/n \geq \alpha$ then $p^{t+1} 2^n \leq p (2p^\alpha)^n$,
which converges to $0$ as $n\rightarrow \infty$.
Thus if we use the sequence of quantum codes described above,
we can make the average of fidelity arbitrary close to $1$
by increasing the code length.
Note that our estimate differs by factor of $2^n$
from an intuition $1 - O(p^{t+1})$ of the fidelity of quantum
error correction.

\subsection{General channel}
The memoryless assumption is used only in Eq.\ (\ref{usememoryless}).
We can calculate a lower bound for the average of the fidelity
over an arbitrary channel as Eq.\ (\ref{beforelast})
by rewriting the unitary operator in a unitary representation
as
\[
\sum_{\vec{a},\vec{b}\in\mathbf{Z}_2^n}
X(\vec{a})Z(\vec{b}) \otimes L_{\vec{a}\vec{b}}.
\]

\subsection{Nonbinary codes}
We can generalize the result to nonbinary stabilizer codes as follows.
We consider $q$-ary stabilizer codes.
Let $H_q$ be the $q$-dimensional Hilbert space and
$\ket{0}$, \ldots, $\ket{q-1}$ an orthonormal basis of $H_q$.
Let $\lambda$ be a primitive $q$-th root of $1$, for example,
$\exp(2\pi i/q)$.
We define a linear map $C_q$ sending $\ket{i}$ to $\ket{i+1 \bmod q}$
and $D_\lambda$ sending $\ket{i}$ to $\lambda^i \ket{i}$ \cite{knill96a}.
Observe that $C_2 = \sigma_x$ and $D_{-1} = \sigma_z$ when
$q=2$.

Let $\Gamma$ be the channel superoperator on $H_q$, and
suppose that a unitary representation of $\Gamma$ is
\[
\Gamma(\rho) = \mathrm{Tr}_{H_E}(U(\rho\otimes |0_E\rangle \langle0_E|)U^*).
\]
We can write $U$ as
\[
U = \sum_{(i,j) \in \mathbf{Z}_q^2} C_q^i D_\lambda^j \otimes L_{i,j},
\]
where $\mathbf{Z}_q = \{0$, \ldots, $q-1\}$ and $L_{i,j}$ is a linear
operator on $H_E$.

Replace the definition of $p$ in Assumption 3 with
\[
p = \sum_{(0,0) \neq (i,j) \in \mathbf{Z}_q^2} \| L_{i,j} 0_E\|^2.
\]
Then the lower bound Eq.\ (\ref{lastevaluation}) also holds for
$q$-ary stabilizer codes.

\subsection{Bounded distance decoding}\label{sec:bounded}
In the error correction process described in
Sec.\  \ref{sec:stabilizer}
we have to find an operator $M \in E$ such that $w(M)$
is minimum among operators $N\in E$ such that $NQ = Q'$.
The task of finding such $M$ from the measurement outcome
becomes computationally difficult when both the code length and
the minimum distance are large \footnote{%
When a stabilizer code is constructed from a $GF(4)$-linear code
with a classical decoding algorithm,
we can use the classical decoding algorithm to determine $M$
from the measurement outcome.
This fact seems a folklore result.
Its nonbinary extension can be found in
Ref.\ \cite[Sec.\  3.2]{matsumotouematsu00}.}.
In practice,
we may give up finding such $M$ if there is no operator $N$
of weight $\leq t'$ such that $NQ = Q'$, where $t'$ is an
integer $\leq t$.
This is a quantum analogue of the classical bounded distance decoding
\cite{macwilliams77}.
We shall slightly modify this bounded distance decoding
and give a lower bound for the average of fidelity.

Let $Q$, $Q'$, $\ket{\psi'}$ and $I_\mathrm{env}$
be as in Sec.\  \ref{sec:stabilizer}.
If there is an operator $N \in E$ such that $NQ = Q'$ and $w(N) \leq t'$,
then let $M=N$. Otherwise choose an operator $M \in E$ such that $MQ = Q'$.
Let the recovered state be $(M^{-1} \otimes I_\mathrm{env}) \ket{\psi'}$.
With this error correction process
the average of the fidelity is bounded from below by
\[
1- \sum_{i=t'+1}^n \left(\begin{array}{c}n\\ i\end{array}\right)
p^i.
\]
The proof is almost the same as that of Eq.\ (\ref{lastevaluation}).

\subsection{Nonstabilizer codes}
It seems difficult to generalize the result in this paper
to nonstabilizer codes.
Because in the error correction of nonstabilizer codes
we cannot write $\ket{\psi}$ as sum of eigenvectors of the
measured observable as in Eq.\ (\ref{sumeigenvector}).

\subsection{Entanglement fidelity}
The entanglement fidelity introduced in Refs.\ \cite{knill97,schumacher96}
should also be considered in some applications,
and
we can estimate the entanglement fidelity from
the fidelity by their relation
\cite[Theorem V.3]{knill97}.

\begin{acknowledgments}
The author would like to thank
the anonymous referee for providing the elaborate report,
Prof.\ Tomohiko Uyematsu for discussion about the inequality in
Sec.\  \ref{sec:asymptotic},
Dr.\ Keiji Matsumoto for pointing out the critical error
in the initial version of this paper,
and Dr.\ Mitsuru Hamada for telling that a similar idea was already used
in Ref.\ \cite[Sec.\  7.4]{preskill99}.
The author was supported by the JSPS Research Fellowship
for Young Scientists during this research.
\end{acknowledgments}


\end{document}